\documentclass{jpsj3}
\usepackage{txfonts}
\usepackage{latexsym,mathrsfs}
\usepackage{bbm}
\usepackage{bm}

\newcommand{\be}{\begin{eqnarray}}
\newcommand{\ee}{\end{eqnarray}}

\newcommand{\Z}{\mathbb{Z}}

\newcommand{\im}{\mathrm{i}}

\newcommand{\mjnal}[4]{#1 {\bf #2} (#4) #3}

\newcommand{\jchp}{J. Chem. Phys.}

\newcommand{\njp}{New J. Phys.}

\newcommand{\pla}{Phys. Lett. A}

\newcommand{\prev}[1][A]{Phys. Rev. #1}

\newcommand{\jmr}{J. Magn. Reson.}
\newcommand{\jpsj}{J. Phys. Soc. Jpn.}
\newcommand{\pnmr}{Prog. NMR Spectrosc.}
\newcommand{\prsocA}{Phil. Trans. R. Soc. Lond. A}

\newcommand{\ve}{\varepsilon}
\newcommand{\vs}{\bm{\sigma}}
\newcommand{\vn}{\bm{n}}
\newcommand{\dvs}{\cdot\vs}
\newcommand{\openone}{I}

\title{Concatenated Composite Pulses Compensating Simultaneous Systematic Errors}

\author{Masamitsu Bando$^1$\thanks{E-mail: bando@alice.math.kindai.ac.jp},
 Tsubasa Ichikawa$^1$\footnote{Present address:
 Department of Physics, Gakushuin University, 1-5-1 Mejiro, Toshima-ku,
 Tokyo 171-8588, Japan},
 Yasushi Kondo$^{1,2}$ and Mikio Nakahara$^{1,2}$}
\inst{$^1$ Research Center for Quantum Computing, Interdisciplinary
 Graduate School of Science and Engineering, Kinki University, 3-4-1
 Kowakae, Higashi-Osaka, Osaka 577-8502, Japan\\
 $^2$ Department of Physics, Kinki University, 3-4-1 Kowakae,
 Higashi-Osaka, Osaka 577-8502, Japan}

\abst{
In NMR experiments and quantum computation, 
many pulse (quantum gate) sequences called the composite pulses,
were developed to suppress one of two 
dominant errors; a pulse length error and an off-resonance error.
We describe, in this paper, a general prescription to design a single-qubit 
concatenated composite pulse (CCCP) that is robust against two types of errors
simultaneously. 
To this end, we introduce a new property, which is satisfied by some 
composite pulses and is sufficient to obtain a CCCP.  
Then we introduce a general method to design 
CCCPs with shorter execution time and less number of pulses. 
}

\kword{composite pulses, quantum gates, NMR}

\begin{document}
\maketitle

\section{Introduction} 

Composite pulses~\cite{counsell85,tycko85,levitt86} 
are important techniques in nuclear magnetic resonance (NMR) experiments 
to suppress systematic errors in simple pulses such as a square pulse, 
which will be called ``elementary pulses'' in the following.
Recently, they receive much attention particularly from a viewpoint of 
quantum computing~\cite{nielsen00,nakahara08}.  
Radio frequency pulses (rf-pulses) in NMR implement quantum gates, 
and composite 
pulses are employed to obtain high precision quantum gates that are robust 
against systematic errors~\cite{jones03,hill07,tomita10,jones11}.  

In NMR, 
there are two typical errors, so called a pulse length error (PLE) and 
an off-resonance error (ORE)~\cite{levitt86}.  
So far, most composite pulses were designed to fight against
one of these two errors; 
BB1~\cite{wim94}, CORPSE and SCROFULOUS~\cite{cumm03}, SK1~\cite{brown04,brown05}, and so on.  
Recently, two new composite pulses were introduced.  
One was due to Alway and Jones~\cite{alway07}, while the other was 
called the Knill pulse~\cite{ryan10}.  
They are robust against these two types of errors simultaneously, although
they are restricted within null operation and $\pi$-rotation (NOT gate) 
of a nuclear spin.  
In order to realize a composite pulse which is robust against these two errors 
simultaneously and without any restriction, we designed 
a ConCatenated Composite Pulse (CCCP) by concatenating CORPSE and 
SCROFULOUS~\cite{ichikawa11}.  

The CCCP reported previously~\cite{ichikawa11} consists of 3 composite pulses, 
each of which consists of 3 elementary pulses. Consequently, this CCCP 
is made of 9 pulses in total. Although this CCCP is robust against 
both PLE and ORE, its execution time is considerably longer than the 
corresponding elementary pulse. CCCPs made of less number of elementary pulses
are certainly desirable from a viewpoint of decoherence suppression.

Establishing general prescription to design a CCCP and
its improvement are the subjects of this paper.  
Some composite pulses have interesting property, which we call the
residual-error-preserving (REP) property. Employing two types of composite 
pulses with mutually exclusive REP properties is essential 
to design a successful CCCP robust against both PLE and ORE. 
By using this method, we obtain many CCCPs systematically.  
Moreover, we further improve the method to reduce operation time 
and the number of elementary pulses of a CCCP.  

In \S\ref{sec:pulse} and \S\ref{sec:cp}, we introduce the
basics of pulses and 
introduce four well known composite pulses, BB1, CORPSE, SCROFULOUS and SK1.  
These composite pulses are robust against either PLE or ORE.  
In \S\ref{sec:rei}, we introduce the REP property to
characterize composite pulses under two types of errors.
Subsequently, we classify composite pulses according to the REP property, 
which is  
an important ingredient to design CCCPs. Then we describe the method to 
design successful CCCPs in \S\ref{sec:cccp}. 
In \S\ref{sec:rcccp}, we show that the number of elementary pulses
to form a CCCP can be reduced if the constituent composite pulse has identity 
elementary pulses as its components. We work out three examples of 
``reduced'' CCCPs.  
The operation times and performances of CCCPs and reduced CCCPs 
are compared in \S\ref{sec:fid}. 
Section~\ref{sec:conc} concludes this paper.

\section{Pulses}\label{sec:pulse}
Consider a two-level system (qubit). A pure state is represented by 
a point on the Bloch sphere and an operation of a single-qubit gate results in 
a rotation of the point around an axis through the centre of the 
Bloch sphere.  

In NMR, we implement any rotations whose axes are in the $xy$-plane 
by controlling timings, strengths and durations of rf-pulses~\cite{claridge}.  
A single-qubit gate without an error in NMR quantum computation
takes the form
\begin{eqnarray}\label{eq:pulse}
 R(\theta,\phi) = \exp[-\im\theta\vn(\phi)\dvs/2],
\end{eqnarray}
where $\theta$ represents the rotation angle, 
$\phi$ is the azimuthal angle which specifies the rotation axis
in the $xy$-plane as 
$\vn(\phi) = (\cos\phi,\sin\phi,0)$ and
$\vs = (\sigma_x,\sigma_y,\sigma_z)$ are the Pauli matrices.
We call $R(\theta,\phi)$ of the form (\ref{eq:pulse}) an elementary pulse, 
following the NMR community convention.

In actual situation, however, errors are unavoidable and 
$R(\theta,\phi)$ is perturbed as 
\begin{eqnarray}
 R^\prime(\theta,\phi) = R(\theta,\phi) + \mathcal{O}(\mathcal{E}), 
\end{eqnarray}
where $\mathcal{E}$ specifies the strength of the error.  
Let $U(\theta,\phi)$ denote a sequence of $N$-pulses as
\begin{eqnarray}
 U(\theta,\phi)
  = R(\theta_N,\phi_N)R(\theta_{N-1},\phi_{N-1})\cdots R(\theta_1,\phi_1),
  \label{eq:comp}
\end{eqnarray}
where the set of elementary pulses
$\{R(\theta_i,\phi_i)\}$ is designed in such a way that 
$U(\theta,\phi)$ reproduces the desired error-free gate $R(\theta,\phi)$
as faithful as possible. Needless to say, $N$ should be taken
as small as possible to avoid decoherence.
Here we assume that the error strengths in all the elementary pulses
are the same, which we will denote by $\mathcal{E}$ as before.
Then eq.~(\ref{eq:comp}) is written as 
\begin{eqnarray}
 U^\prime(\theta,\phi)
 = R(\theta,\phi) - \im\mathcal{E}\delta U + \mathcal{O}(\mathcal{E}^2),
\end{eqnarray}
where $U^\prime(\theta,\phi)$ represents $U(\theta,\phi)$ in the presence of 
error and $\delta U$ gives the structure of the first order error term.  
The set of elementary gates $\{R(\theta_i,\phi_i)\}$ is called a
composite pulse if it is arranged to make $\delta U$ vanish so that
\begin{eqnarray}
 U^\prime(\theta,\phi) = R(\theta,\phi) + \mathcal{O}(\mathcal{E}^2).
\end{eqnarray}

\section{Composite Pulses in NMR}\label{sec:cp}
In this section, we review typical composite pulses that are robust 
against one of two systematic errors in NMR: 
one is a PLE and the other is an ORE.  
For simplicity we henceforth ignore the second and higher order error terms.  

\subsection{Pulse length error}

When a PLE is present, an elementary pulse $R(\theta,\phi)$ turns to
\begin{eqnarray}
 R^\prime_{\ve}(\theta,\phi) = R((1+\ve)\theta,\phi)
 \approx R(\theta,\phi) -\im\ve\theta\; (\vn(\phi)\dvs)\; R(\theta,\phi)/2,\label{eq:pl}
\end{eqnarray}
where $R^\prime_\ve (\theta,\phi)$ is the actual pulse in the presence
of the PLE and $\ve$ is an unknown but a fixed constant that 
represents the strength of the PLE. 
Some composite pulses robust against PLE are well-known in the NMR
community:  
BB1~\cite{wim94}, SCROFULOUS~\cite{cumm03} and SK1~\cite{brown04}, to name
a few.
The BB1 consists of four elementary pulses with parameters
\begin{eqnarray}
 \theta_1 = \theta_3 = \pi,\quad \theta_2 = 2\pi, \quad \theta_4 = \theta,\nonumber\\
 \phi_1 = \phi_3 = \phi + \arccos[-\theta/(4\pi)], \quad
  \phi_2 = 3\phi_1 - 2\phi, \quad \phi_4 = \phi, \label{eq:bb1}
\end{eqnarray}
while the SCROFULOUS consists of three elementary pulses with parameters
\begin{eqnarray}
  \theta_1 = \theta_3 = \mathrm{arcsinc}[2\cos(\theta/2)/\pi], \quad
   \theta_2 = \pi, \nonumber\\
  \phi_1 = \phi_3 = \arccos[-\pi\cos\theta_1/(2\theta_1\sin(\theta/2))],\nonumber\\
  \phi_2 = \phi_1 - \arccos[-\pi/(2\theta_1)],
\end{eqnarray}
where $\mathrm{sinc}\,\theta = \sin\theta/\theta$.  
The SK1 also consists of three elementary pulses with parameters
\begin{eqnarray}
 \theta_1 = \theta,\quad \theta_2 = \theta_3 = 2\pi,\nonumber\\
 \phi_1 = \phi, \quad \phi_2 = \phi - \arccos[-\theta/(4\pi)], \quad
  \phi_3 = \phi + \arccos[-\theta/(4\pi)]. \label{eq:sk1}
\end{eqnarray}

To diminish the effects of random noises, 
geometric quantum gates~\cite{zanardi99,zhu02,ota09-1} 
based on the holonomy~\cite{berry84,wilczek84,aharonov87,nakahara03}
associated with the geometrical setting of the system
have been proposed.  
All composite pulses that are robust against PLE are 
found to be geometric quantum gates~\cite{kondo11,pt2012}.

\subsection{Off-resonance error}

When an ORE is present, an elementary pulse 
$R(\theta,\phi)$ turns to 
\begin{eqnarray}
 R^\prime_{f}(\theta,\phi)
  = \exp[-\im\theta(\vn(\phi)\dvs + f\sigma_z)/2]
  \approx R(\theta,\phi) -\im f\sin(\theta/2)\sigma_z, \label{eq:or}
\end{eqnarray}
where $R^\prime_f (\theta,\phi)$ is the actual elementary pulse
in the presence of the ORE and $f$ is an unknown constant that 
characterizes the strength of the ORE. The  
CORPSE pulse sequence 
is the best known composite pulse to suppress ORE~\cite{cumm03}. 
CORPSE consists of three elementary pulses 
\begin{eqnarray}
 \theta_1 = 2n_1\pi + \theta/2 - k, \quad \theta_2 = 2n_2\pi - 2k, \quad
  \theta_3 = 2n_3\pi + \theta/2 - k,\nonumber\\
 \phi_1 = \phi_2 - \pi = \phi_3 = \phi,
\end{eqnarray}
where $k = \arcsin[\sin(\theta/2)/2]$ and $n_i\in\Z\ (i = 1,2,3)$.  
It is common to take $n_1 = n_2 = 1$ and $n_3 = 0$ for CORPSE
and the one with $n_1 = n_3 = 0$, $n_2 = 1$ is called the short CORPSE.

\subsection{Triviality of $N=2$ composite pulses}\label{no-go}

We first define a pulse is trivial, regardress whether it is elementary or
composite, when the resulting operation is the identity operation up to the
overall phase.
We now prove that there are no non-trivial $N = 2$ composite pulses.  
We need at least $N = 3$ in order to implement 
non-trivial composite pulses in a robust way. 
(It was shown in \cite{2qubit} that this is also the case for a
two-qubit composite pulse robust against a $J$-coupling error.)

Consider a composite pulse with $N=2$ 
\begin{eqnarray}
 U(\theta,\phi) = R(\theta_2,\phi_2)R(\theta_1,\phi_1). \label{eq:cp-n2}
\end{eqnarray}
First, suppose that this $U(\theta, \phi)$ is robust against 
a PLE.  
Equation~(\ref{eq:cp-n2}) in the presence of PLE is written as 
\begin{eqnarray}
 U^\prime(\theta,\phi)
  \approx R(\theta,\phi) -\im\ve R(\theta_2,\phi_2)
 \left[ (\theta_1\vn(\phi_1) + \theta_2\vn(\phi_2))\dvs\right]
 R(\theta_1,\phi_1)/2,
\end{eqnarray}
where $R(\theta,\phi) = R(\theta_2,\phi_2)R(\theta_1,\phi_1)$.
The robustness condition requires that the first order error term must
vanish, which leads to
\begin{eqnarray}
 \theta_1\vn(\phi_1)\dvs = - \theta_2\vn(\phi_2)\dvs.
\end{eqnarray}
This condition is satisfied if we make
the following choice: 
\begin{eqnarray}
 \theta_2 = -\theta_1,\quad \phi_2 = \phi_1 \quad {\rm or} \quad
  \theta_2 = \theta_1,\quad \phi_2 = \phi_1 + \pi. \label{eq:n2-pl-cond}
\end{eqnarray}
Substituting eq.~(\ref{eq:n2-pl-cond}) to eq.~(\ref{eq:cp-n2}), 
we find $U(\theta,\phi)=\openone$, where $\openone$ is the $2\times 2$ 
identity operator.  
This proves triviality of any $N=2$ composite pulse
robust against PLE.

Next, suppose that the composite pulse (\ref{eq:cp-n2}) is robust against the 
ORE. A composite pulse under ORE is written as 
\begin{eqnarray}
 U^\prime(\theta,\phi) = R(\theta,\phi) -\im f\left(\sin(\theta_1/2)R(\theta_2,\phi_2)
 + \sin(\theta_2/2)R^\dag(\theta_1,\phi_1)\right)\sigma_z. 
\end{eqnarray}
Then the robustness condition is found to be
\begin{eqnarray}
 \sin(\theta_1/2)R(\theta,\phi) + \sin(\theta_2/2)\openone = 0, \label{eq:cond_a2}
\end{eqnarray}
which is satisfied if $\theta=2\pi n\ (n\in\Z)$. This proves that
an $N=2$ composite pulse robust against ORE is trivial.

These two observations reveal that a non-trivial composite pulse 
robust against either PLE or ORE requires 
three elementary pulses or more.  


\section{Residual-Error-Preserving Properties}\label{sec:rei}
In this section, we introduce an important property of composite pulses 
that we call the Residual-Error-Preserving (REP) property. 
Consider a case in which both PLE and ORE are present. 
Then an elementary pulse is perturbed as
\begin{eqnarray}
 R^\prime(\theta,\phi)
  = \exp[-\im (1+\ve)\theta (\vn(\phi)\dvs + f\sigma_z)/2].\label{eq:pl-unw}
\end{eqnarray}
Let us introduce  
$\mathcal{R}(\theta,\phi,\delta_\ve U, \delta_f U)$, which is 
an elementary pulse taking into account two errors to the first order as
\begin{eqnarray}
 \mathcal{R}(\theta,\phi,\delta_\ve U,\delta_f U)
  =
 R(\theta,\phi) -\im\ve \delta_\ve U\; (\vn(\phi)\dvs )\;R(\theta,\phi)/2
  -\im f \delta_f U\sigma_z, \label{eq:w}
\end{eqnarray}
where 
$\delta U_\ve$ and $\delta U_f$ are matrices, in general,
characterizing the
first order error terms for PLE and ORE, respectively.  
According to this notation, the elementary pulse (\ref{eq:pl-unw}) is
rewritten, to the first order, as
\begin{eqnarray}
 R^\prime(\theta,\phi) = \mathcal{R}(\theta,\phi,\theta,\sin(\theta/2)).
  \label{eq:pl-and-or}
\end{eqnarray}
Note that both $\delta_{\ve} U$ and $\delta_f U$
reduce to scalars $\theta$ and $\sin \theta/2$, respectively,
for 
an elementary gate.  

Similarly, CORPSE is written in terms of $\mathcal{R}$ as
\begin{eqnarray}
 U^\prime_{\rm CORPSE}(\theta,\phi)
  &=& R^\prime(\theta_3,\phi)R^\prime(\theta_2,\bar{\phi})R^\prime(\theta_1,\phi)
  \nonumber\\
 &=& R(\theta,\phi) - \im\ve(\theta_1 - \theta_2 + \theta_3)\vn(\phi)\dvs
  R(\theta,\phi)/2\nonumber\\
 &=& R(\theta,\phi) - \im\ve\theta\vn(\phi)\dvs R(\theta,\phi)/2\nonumber\\
 &=& \mathcal{R}(\theta,\phi,\theta,0),\label{eq:rei-corpse}
\end{eqnarray}
where the defining relations $\bar{\phi} = \phi+\pi$ and
$\theta_1 - \theta_2 + \theta_3 = \theta$ of CORPSE have been used. 
The last line of the above equation shows that
$U^\prime_{\rm CORPSE}(\theta,\phi)$
is regarded as the target elementary pulse under the influence of the PLE only
(See eq.~(\ref{eq:pl})), eliminating the effect of ORE.
Similarly, $U^\prime_{\rm SK1}(\theta,\phi)$ and $U^\prime_{\rm BB1}
(\theta,\phi)$, the SK1 and BB1 pulses, respectively, in the presence of
two errors reduce to
\begin{eqnarray}
 U^\prime_{\rm SK1}(\theta,\phi) = U^\prime_{\rm BB1}(\theta,\phi)
  \approx \mathcal{R}(\theta,\phi,0,\sin(\theta/2)), \label{eq:rei-sk-bb1}
\end{eqnarray}
namely a pulse with the ORE only.
In contrast, $U^\prime_{\rm SCROF}(\theta,\phi)$, 
the SCROFULOUS pulse to the first order, reduces to 
\begin{eqnarray}
 U^\prime_{\rm SCROF}(\theta,\phi)
  = \mathcal{R}(\theta,\phi,0,\delta_f U), \label{eq:rei-scrof}
\end{eqnarray}
where
\begin{eqnarray}
 \delta_f U
  &=& \left[\sin\theta_1\cot(\theta_2/2)
       - 2\sin^2(\theta_1/2)(\cos(\phi_1 - \phi_2))
       + 1\right]\sin(\theta_2/2),
\end{eqnarray}
which is different from
$\delta_f U = \sin (\theta/2)$ of an elementary pulse.  
SCROFULOUS behaves differently
from an elementary pulse in this respect. This apparently minor difference
plays an essential role in designing a composite pulse robust against two
types of errors simultaneously.

It is found from 
eqs.~(\ref{eq:pl-and-or}), (\ref{eq:rei-corpse}) and 
(\ref{eq:rei-sk-bb1}) that the first order PLE error term 
$\theta$ of 
CORPSE is the same as that of the target elementary 
pulse and CORPSE is regarded as an elementary pulse under PLE only,
while the first order 
ORE error terms $\sin \theta/2$ of SK1 and BB1 are 
the same as those of the target elementary pulses
and SK1 and BB1 are regarded as elementary pulses under ORE only.
We call these properties ``Residual-Error-Preserving'' (REP). 

With these observations, we introduce the following definitions.
\begin{itemize}
\item A composite pulse $U(\theta, \phi)$ robust against PLE satisfying 
the property
\begin{equation}
U^\prime(\theta,\phi)=\mathcal{R}(\theta,\phi,0,\sin(\theta/2))
\end{equation}
is called residual-error-preserving with respect to ORE (REP-ORE).
\item A composite pulse $U(\theta, \phi)$ robust against ORE satisfying
the property
\begin{equation}
U^\prime(\theta,\phi)=
\mathcal{R}(\theta,\phi,\theta,0)
\end{equation}
is called residual-error-preserving with respect to PLE (REP-PLE).
\end{itemize}

For example, CORPSE is REP-PLE, while
SK1 and BB1 are REP-ORE.
In contrast, SCROFULOUS does not have the REP property.  
Types of REP and robustness of well known composite pulses are summarized in 
Table~\ref{tab:rei}.  
The existence of REP property is important to design concatenated
composite pulses as described in the next section.  

\begin{table}[t]
 \caption{Residual-error-preserving (REP) property
and robustness of composite pulses.  
 The entry ``REP'' shows the type of REP property while 
 the entry ``robustness'' shows the error type against which
 the composite pulse is robust.
 SK1, BB1 and CORPSE can be employed as an inner composite pulse,
 to be defined in the next section, in CCCP since they are REP.}
 \label{tab:rei}
 \begin{center}
  \begin{tabular}{l@{\hspace{10mm}}c@{\hspace{10mm}}c}\hline
   composite pulse                    & REP & robustness\\\hline\hline
   SK1                                & ORE & PLE\\
   BB1                                & ORE & PLE\\
   SCROFULOUS                         & --  & PLE\\
   CORPSE ($n_1=n_2=1,\ n_3=0$)       & PLE & ORE\\
   short CORPSE ($n_1=n_3=0,\ n_2=1$) & --  & ORE\\\hline
  \end{tabular}
 \end{center}
\end{table}

\section{Concatenated Composite Pulses}\label{sec:cccp}

We now show how to design a concatenated composite pulse (CCCP).  
By concatenating two different composite pulses robust against two different 
types of errors, we obtain various CCCPs robust against both types of errors 
simultaneously~\cite{ichikawa11}.

\begin{figure}[t]
 \center\includegraphics[clip,scale=0.8]{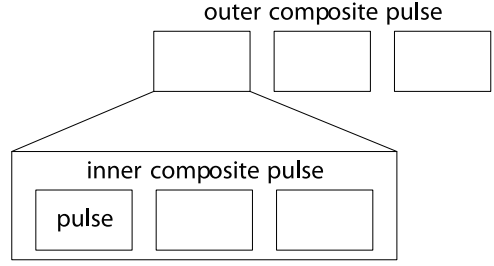}
 \caption{Schematic diagram of outer composite pulse and inner composite 
 pulse.  
 Each elementary pulse in the outer composite pulse is replaced by the
 inner composite pulse. Time goes from left to right in this
 and the following pulse diagrams.}
 \label{fig:cccp}
\end{figure}

Suppose there is a composite pulse robust against PLE (ORE).
Now it should be clear that we need to
replace its constituent elementary
pulses by other composite pulses which are 
robust against ORE (PLE) and REP-PLE (REP-ORE) to design a CCCP
robust against PLE and ORE simultaneously.
We call the former composite pulse {\it outer} while the latter one
 {\it inner}. Figure~\ref{fig:cccp} explains this naming convention.

Let us confirm the above statement by explicitly examining the pulses.
Let $V(\theta, \phi)=R(\theta_N, \phi_N) \ldots R(\theta_1,\phi_1)$
be an outer composite pulse robust against PLE. Each elementary 
pulse $R(\theta_i, \phi_i)$ is replaced by an inner composite pulse 
$U(\theta_i, \phi_i)$ robust against ORE and are REP-PLE following
the above prescription.
The CCCP now takes the form $V(\theta, \phi)
= U(\theta_N, \phi_N) \ldots U(\theta_1,\phi_1)$. Under both PLE and
ORE, the inner composite pulse is perturbed as
$U'(\theta_i, \phi_i) = {\mathcal{R}}(\theta_i, \phi_i, \theta_i, 0)$
by definition. This CCCP is robust against both types of errors since
\begin{eqnarray*}
V'(\theta, \phi)
&= &U'(\theta_N, \phi_N) \ldots U'(\theta_1,\phi_1)\\
&=&  {\mathcal{R}}(\theta_N, \phi_N, \theta_N, 0) \ldots 
{\mathcal{R}}(\theta_1, \phi_1, \theta_1, 0) \\
&=&  {\mathcal{R}}(\theta, \phi, 0, 0).
\end{eqnarray*}
We have used the fact that $V$ is a composite pulse robust 
against PLE to derive the last equality.  

Various CCCPs are obtained by choosing the outer and 
the inner composite pulses according to this scheme as listed in 
Table~\ref{tab:cccp}.  
It was shown in \S\ref{sec:cp} that 
a non-trivial composite pulse robust against either 
the PLE or the ORE requires at least three pulses. 
Therefore the method to design a CCCP introduced here
requires at least $N\ge 9$ pulses to implement a non-trivial CCCP.

\begin{table}[t]
 \caption{CCCPs obtained with our scheme and their
 inner and outer composite pulses.  Note that the outer composite pulses of
 SKinsC and BBinsC are the short CORPSE.}
 \label{tab:cccp}
 \begin{center}
  \begin{tabular}{l@{\hspace{12mm}}c@{\hspace{12mm}}c}\hline
                        & \multicolumn{2}{c}{composite pulse}\\
   abbreviation of CCCP & inner  & outer\\\hline\hline
   CinS                 & CORPSE & SCROFULOUS\\
   CinSK                & CORPSE & SK1\\
   SKinsC               & SK1    & short CORPSE\\
   CinBB                & CORPSE & BB1\\
   BBinsC               & BB1    & short CORPSE\\\hline
  \end{tabular}
 \end{center}
\end{table}

\section{Reduced Concatenated Composite Pulses}\label{sec:rcccp}
In this section, 
we employ composite pulses having trivial elementary pulses as constituents
to reduce the number of elementary pulses in the resulting CCCP.
Note that a trivial composite pulse robust against PLE or ORE 
can be constructed with only two elementary pulses and does not require 
three elementary pulses like in the case of non-trivial composite pulses
as was proved in \S\ref{sec:cp}.  

We use three trivial (composite) pulses to this end in the following.  
Let us first consider $R(\theta, \phi)R(\theta, \bar{\phi})$, 
which reduces to the identity operator when no errors are present.  
From eq.~(\ref{eq:pl}), the pulse sequence 
$R^\prime(\theta, \phi) R^\prime(\theta, \bar{\phi})$ under both PLE and ORE
is 
\begin{eqnarray}
 && R^\prime(\theta, \phi)R^\prime(\theta,\bar{\phi})\nonumber\\
 && = R(0,\phi) - \im\ve(\theta - \theta)\vn(\phi)\dvs R(0,\phi)/2
  -2\im f\sin(\theta/2)R(\theta,\phi)\sigma_z\nonumber\\
 && = \mathcal{R}(0,\phi,0,2\sin(\theta/2)R(\theta,\phi)), \label{eq:triv-pl}
\end{eqnarray}
showing that the trivial pulse sequence $R(\theta, \phi)R(\theta, \bar{\phi})$ 
is robust against PLE.  

An ideal pulse $R(2\pi,\phi)$ and a pulse sequence 
$R(\pi,\phi^\prime)R(2\pi,\phi)R(\pi,\phi^\prime)$ are trivial pulses 
up to the overall phase. 
These pulses under PLE and ORE are 
\begin{eqnarray}
 R^\prime(2\pi,\phi) &=&  -\openone + \im\ve\pi\vn(\phi)\dvs
  - \im f\sin(\pi)\sigma_z\nonumber\\
 &=& -\mathcal{R}(0,\phi,2\pi, 0) ,\quad\label{eq:triv-or-1}
\end{eqnarray}
and
\begin{eqnarray}
 && R^\prime(\pi,\phi^\prime)R^\prime(2\pi, \phi)
  R^\prime(\pi,\phi^\prime)\nonumber\\
 && = \openone -\im\ve\pi\left( \vn(\phi^\prime) + \vn(2\phi^\prime -\phi)
                         \right)\dvs + \im f[(\sigma_z R(\pi,\phi^\prime)
 + R(\pi,\phi^\prime)\sigma_z)]\nonumber\\
  &&= \mathcal{R}(0,\phi,\delta_\ve U,0),\label{eq:triv-or-2}
\end{eqnarray}
where
\begin{eqnarray}
 \delta_\ve U = 2\pi\left[
  \exp(-\im(\phi^\prime -\phi)\sigma_z) + 
  \exp(-2\im(\phi^\prime -\phi)\sigma_z)\right], 
\end{eqnarray}
and we have used $\sigma_z R(\pi,\phi^\prime) = - R(\pi,\phi^\prime)\sigma_z$.  
Therefore these pulses are robust against the ORE.  
Table~\ref{tab:triv} summarizes robustness of these three trivial
pulses. Although it might seem that the $N=3$ trivial composite pulse 
is useless, it has an important use in reduction of elementary pulses
in a CCCP as we show in the following examples.

\begin{table}[t]
\caption{Robustness of three trivial composite pulses. 
 Here $\bar{\phi} = \phi + \pi$. 
 They are robust against either the PLE or the ORE.}
 \label{tab:triv}
 \begin{center}
  \begin{tabular}{l@{\hspace{40mm}}c}\hline
   pulse sequence & robustness\\\hline\hline
   $R(\theta,\phi)R(\theta,\bar{\phi})$ & PLE\\
   $R(2\pi,\phi)$                       & ORE\\
   $R(\pi,\phi^\prime)R(2\pi,\phi)R(\pi,\phi^\prime)$ & ORE\\\hline
  \end{tabular}
 \end{center}
\end{table}

As far as we use concatenation, 
a composite pulse that is robust against both PLE and ORE 
must be composed of at least 5 pulses.  
As discussed in \S\ref{sec:cp}, 
the minimum number of components for the outer and inner 
composite pulses is three.  
And the outer composite pulse cannot be composed of trivial pulses only.  
Then we must have at least one non-trivial pulse as a constituent
for the outer composite pulse (SK1 is an example).  
This pulse must be replaced by an inner composite pulse.  
For the trivial composite pulses that are robust against either PLE or ORE, 
there are options of $N=1,2$ and $3$ as given in Table~\ref{tab:triv}.  
This proves that the outer composite pulse can contain one or more 
trivial composite pulses, and so the minimum $N$ for a CCCP is $1+1+3 = 5$.  
We show three convincing examples of reduced CCCPs below.

\subsection{Example 1}
\begin{figure}
 \center\includegraphics[clip,scale=0.8]{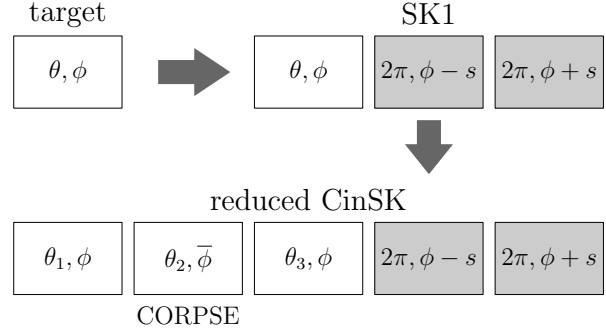}
 \caption{Schematic diagram to design the reduced CinSK.  
 Each box represents a pulse and is parameterized by the rotation angle and 
 the azimuthal angle of the rotation axis. Here 
 $s=\arccos[-\theta/(4\pi)]$ and $\bar{\phi} = \phi + \pi$.  
 Shaded boxes are pulses requiring no replacements by composite
 pulses.}
 \label{fig:rcisk-cccp}
\end{figure}

The first example is the reduced CinSK.  
First, we replace a target pulse $R(\theta,\phi)$ by the SK1~(\ref{eq:sk1}).  
Next, we select the CORPSE as the inner composite pulse because it is robust 
against the ORE and REP-PLE.  
Since the SK1 is robust against PLE and has two trivial pulses 
$R(2\pi, \phi \pm \arccos[-\theta/(4\pi)])$ 
as the second and the third pulses,  
we need to replace only the first pulse by CORPSE,
remembering that these
trivial pulses are already robust against ORE.
Therefore, we obtain a reduced CORPSE in SK1 CCCP (reduced CinSK) with 
$N = 3 + 1 + 1 = 5$ pulses.  
The reduced CinSK is parameterized as
\begin{eqnarray}
 \theta_1 = \theta_3 + 2\pi = 2\pi + \theta/2 - k, \quad
  \theta_2 = 2\pi - 2k,\quad \theta_4 = \theta_5 = 2\pi,\nonumber\\
 \phi_1 = \phi_2 - \pi = \phi_3 = \phi,\nonumber\\
 \phi_4 = \phi-\arccos[-\theta/(4\pi)], \quad
  \phi_5 = \phi+\arccos[-\theta/(4\pi)].
 \end{eqnarray}
Schematic diagram to design the reduced CinSK is shown in 
Fig.~\ref{fig:rcisk-cccp}.  

\subsection{Example 2}
\begin{figure}
\center\includegraphics[clip,scale=0.8]{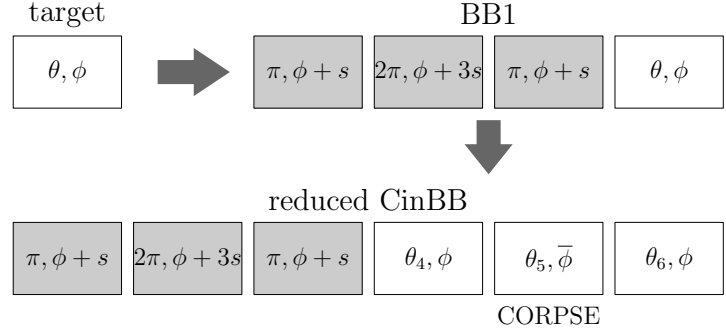}
 \caption{Schematic diagrams to design the reduced CinBB.  
 Each box represents a pulse and is parameterized by the rotation angles and 
 azimuthal angle of the rotation axis, $s=\arccos[-\theta/(4\pi)]$ and 
 $\bar{\phi} = \phi + \pi$. Shaded boxes are pulses requiring no 
 replacements by composite pulses.}
 \label{fig:rcibb-cccp}
\end{figure}

The second example is the reduced CinBB.  
First, we replace a target pulse $R(\theta,\phi)$ by BB1~(\ref{eq:bb1}), 
which is robust against PLE.
Next, we select CORPSE as the inner composite pulse since 
it is robust against ORE and REP-PLE, which guarantees a successful CCCP.
Since BB1 is robust against PLE and contains a trivial pulse sequence 
$R(\pi, \phi_1)R(2\pi, \phi_2)R(\pi, \phi_1)$ that is robust against ORE, 
we need to replace only the fourth pulse in BB1 by CORPSE.  
As a result, we obtain the reduced CORPSE in BB1 CCCP (reduced CinBB) with 
$N = 1 + 1 + 1 + 3 = 6$ elementary pulses.  
The reduced CinBB is parameterized as 
\begin{eqnarray}
 \theta_1 = \theta_3 = \pi, \quad \theta_2 = 2\pi,\quad \theta_4 = \theta_6 + 2\pi = 2\pi + \theta/2 - k,\nonumber\\
 \theta_5 = 2\pi - 2k,\quad
 \phi_1 = \phi_3 = \phi + \arccos[-\theta/(4\pi)],\nonumber\\
 \phi_2 = 3\phi_1 - 2\phi,\quad \phi_4 = \phi_5 - \pi = \phi_6 = \phi.
\end{eqnarray}
Schematic diagram of the reduced CinBB is shown in 
Fig.~\ref{fig:rcibb-cccp}.  

\subsection{Example 3}
The last example is the reduced SKinsC.  
First, we replace a 
target pulse $R(\theta,\phi)$ by the short CORPSE robust against ORE.  
While the short CORPSE has no trivial pulse sequences, 
it is modified in order to include trivial pulse sequences.  
For this, here we use an important property of a pulse sequence 
$\prod_{i=1}^{N}R(\theta_i,\phi)$ with $\sum_{i=1}^N\theta_i = \theta$.  
This pulse sequence is equivalent to $R(\theta,\phi)$ in the error-free case, 
and from eq.~(\ref{eq:pl-unw}), 
this pulse sequence under the PLE and the ORE becomes
\begin{eqnarray}
 \prod_{i=1}^{N}R^\prime(\theta_i,\phi)
  &=& R^\prime(\theta,\phi).\label{eq:commute}
\end{eqnarray}
Note that 
$R^\prime(\theta_i,\phi)R^\prime(\theta_j,\phi) = R^\prime(\theta_i+\theta_j,\phi)$ 
for $i,j=1,2,\ldots,N$ due to the commutativity 
$[R^\prime(\theta_i,\phi), R^\prime(\theta_j,\phi)] = 0$.  
This shows the pulse sequence $\prod_{i=1}^{N}R(\theta_i,\phi)$ is
both REP-PLE and REP-ORE.  
By using this result, 
we can modify the short CORPSE for the target elementary
pulse $R(\theta,\phi)$ as follows: 
\begin{eqnarray}
 \lefteqn{R(\theta_3,\phi) R(\theta_2,\bar\phi) R(\theta_1,\phi)}\nonumber\\
 &=& R(\theta_3,\phi) R(\theta_3,\bar\phi) 
  R(2\pi-\theta,\bar\phi)  R(\theta_1,\bar\phi) R(\theta_1,\phi),
  \label{eq:scorpse}
\end{eqnarray}
where $\theta_1 = \theta_3 = \theta/2 - k$, $\theta_2 = 2\pi - 2k$, 
and $k = \arcsin[\sin(\theta/2)/2]$.  
Two trivial pulse sequences $R(\theta_1,\bar\phi)R(\theta_1,\phi)$ and 
$R(\theta_3,\phi)R(\theta_3,\bar\phi)$ in the modified short CORPSE are 
robust against  PLE as listed in Table~\ref{tab:triv}.  

\begin{figure}
 \center\includegraphics[clip,scale=1.0]{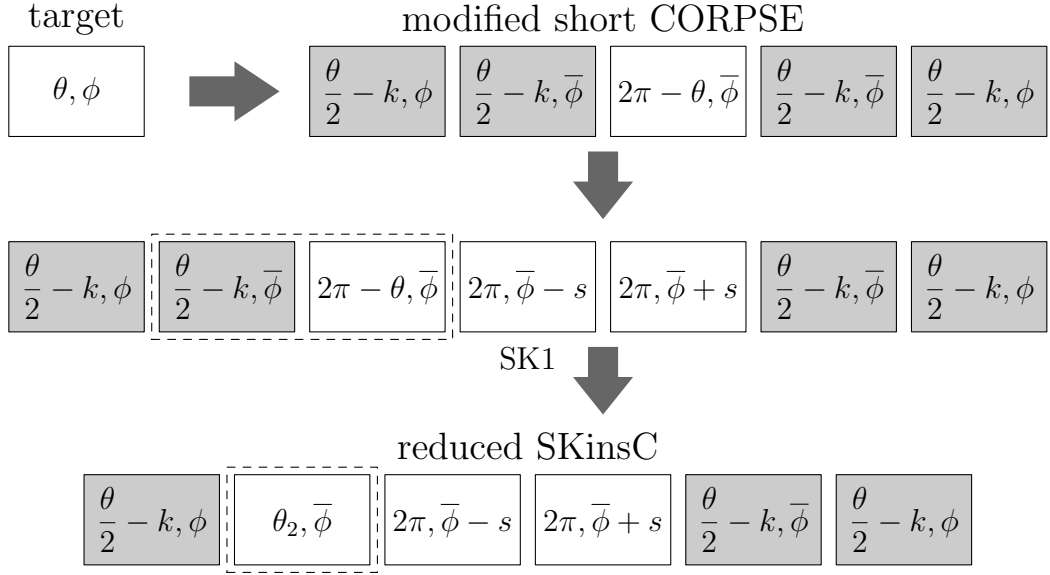}\\
 \caption{Schematic diagram to design the reduced SKinsC.  
 Each box represents a pulse and is parameterized by its rotation angle and 
 the azimuthal angle of the rotation axis. Here $s=\arccos[-\theta/(4\pi)]$, 
 $k=\arcsin[\sin(\theta/2)/2]$ and $\bar{\phi} = \phi + \pi$.  
 Shaded boxes are pulses requiring no replacements by composite pulses.  
 The second and the third pulses in the middle row are merged 
 in the bottom row (surrounded by dashed boxes).  }
 \label{fig:rskic-cccp}
\end{figure}

Next, we choose SK1 as an inner composite pulse since it is robust against 
PLE and REP-ORE.  
Since the modified short CORPSE has two trivial pulse sequences 
$R(\theta_1,\bar\phi) R(\theta_1,\phi)$ and 
$R(\theta_3,\phi) R(\theta_3,\bar\phi)$ robust against PLE, 
we need to replace only the third pulse by SK1.  
In addition, by using eq.~(\ref{eq:commute}), we can merge the second 
and the third pulses without affecting the robustness issue 
since $\phi_2 = \phi_3$.  
Therefore, we obtain the reduced SK1 in modified short CORPSE CCCP 
(reduced SKinsC) 
which has $N = 2 + 3 + 2 - 1 = 6$ pulses. The term $-1$ in $N$ accounts 
for the
reduction of pulses by merging two pulses. The reduced SKinsC is 
parameterized as
\begin{eqnarray}
 \theta_1 = \theta_5 = \theta_6 = \theta/2 - k, \quad
  \theta_2 = 2\pi - \theta/2 - k, \quad \theta_3 = \theta_4 = 2\pi,\nonumber\\
 \phi_1 = \phi_2 - \pi = \phi_5 - \pi = \phi_6 = \phi,\nonumber\\
 \phi_3 = \phi-\arccos[-\theta/(4\pi)], \quad
  \phi_4 = \phi+\arccos[-\theta/(4\pi)]. 
\end{eqnarray}
Schematic diagram to design the reduced SKinsC is shown in 
Fig.~\ref{fig:rskic-cccp}.  

\section{Operation Time Cost and Fidelity}\label{sec:fid}
\begin{table}[t]
 \caption{Number of pulses, $N$, operation time
 cost $T$ and robustness of the elementary target pulse,
 composite pulses and CCCPs.  
 $T(\pi/2)$ and $T(\pi)$ are the operation time cost $T$ with the
 target rotation angle $\pi/2$ and $\pi$, respectively.  
 The ``robustness'' represents the error type against which the pulse is 
 robust.
 Reduced CinBB and SKinsC attain approximately
 50\% reduction in the operation time cost compared to their nonreduced 
 counterparts.}
 \label{tab:time-avgf}
 \begin{center}
  \begin{tabular}{l@{\hspace{8mm}}c@{\hspace{8mm}}c@{\hspace{8mm}}c@{\hspace{8mm}}c@{\hspace{2mm}}c}\hline
   pulse           & $N$  & $T(\pi/2)$ & $T(\pi)$ & robustness\\\hline\hline
   elementary      & $1$  & $0.5$      & $1.0$    & -- \\\hline
   SCROFULOUS      & $3$  & $2.3$      & $3.0$    & PLE \\
   SK1             & $3$  & $4.5$      & $5.0$    & PLE \\
   BB1             & $4$  & $4.5$      & $5.0$    & PLE \\
   short CORPSE    & $3$  & $2.0$      & $2.3$    & ORE \\
   CORPSE          & $3$  & $4.0$      & $4.3$    & ORE \\\hline
   CinS            & $9$  & $12.5$     & $13.0$   & PLE, ORE \\
   CinSK           & $9$  & $16.0$     & $16.3$   & PLE, ORE \\
   CinBB           & $12$ & $18.7$     & $19.0$   & PLE, ORE \\
   SKinsC          & $9$  & $14.0$     & $14.3$   & PLE, ORE \\
   BBinsC          & $12$ & $14.0$     & $14.3$   & PLE, ORE \\\hline
   reduced CinSK   & $5$  & $8.0$      & $8.3$    & PLE, ORE \\
   reduced CinBB   & $6$  & $8.0$      & $8.3$    & PLE, ORE \\
   reduced SKinsC  & $6$  & $6.0$      & $6.3$    & PLE, ORE \\\hline
  \end{tabular}
 \end{center}
\end{table}

A reduced CCCP has a shorter operation time than the original CCCP.  
In addition, generally, it has improved robustness thanks to
the short operation time and less number of elementary pulses. 
We introduce two measures,
the operation time cost $T$ and the gate (pulse) fidelity $F$,
to compare performances of CCCPs.

The operation time cost $T$ is defined by 
\begin{eqnarray}
  T = \sum_{i=1}^{N} \theta_i/\pi,
\end{eqnarray}
where $\theta_i$ is a rotation angle of the $i$-th pulse.  

As an example, let us consider CinSK parameterized as
\begin{eqnarray}
 \theta_1 &=& \theta_3 + 2\pi = 2\pi + \theta/2 - k,\quad
  \theta_2 = 2\pi -2k,\nonumber\\
 \theta_4 &=& \theta_7 = 3\pi,\quad \theta_6 = \theta_9 = \pi,\quad 
  \theta_5 = \theta_8 = 2\pi. 
\end{eqnarray}
The operation time cost of CinSK is easily found as
\begin{eqnarray}
 T = 16 + (\theta - 4k)/\pi.
\end{eqnarray}
This should be compared with that of the reduced CinSK, 
\begin{eqnarray}
 T = 8 + (\theta - 4k)/\pi.
\end{eqnarray}
Number of elementary pulses, $N$, and the numerical values of operation time 
cost $T$ for the elementary pulse, some composite pulses and CCCPs with the
angle 
$\theta=\pi/2$ and $\pi$ are listed in Table~\ref{tab:time-avgf}.  
It shows that a reduced CCCP achieves approximately 50\% reduction in the
operation time cost compared to its nonreduced counterpart.

\begin{figure}[t]
 \center\includegraphics[clip,scale=0.8]{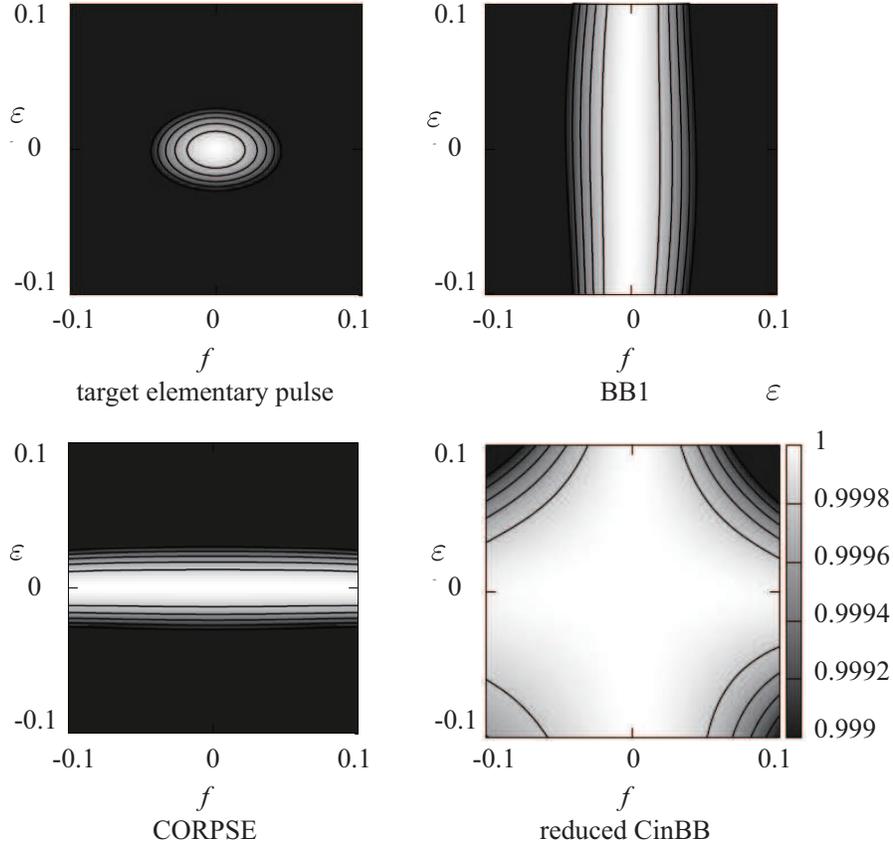}
 \caption{Fidelity $F$ of a target elementary pulse, BB1, CORPSE, and 
 reduced CinBB as a 
 function of the error strengths $\ve$ for PLE and $f$ 
 for ORE.  The target rotation angle is $\theta=\pi$.  
 The reduced CinBB is robust against PLE and ORE simultaneously.  }
 \label{fig:dfplot}
\end{figure}

The gate fidelity of $U^\prime$ is defined by 
\begin{eqnarray}
 F = |\tr(U^\dag U^\prime)|/2,
\end{eqnarray}
where $U$ is an ideal pulse without errors corresponding $U^\prime$.  
The gate fidelity $F$ is a commonly used measure of a quantum gate accuracy 
based on the Hilbert-Schmidt inner product with respect to 
the one-qubit Hilbert space~\cite{nielsen00,alway07,jones11,hill07}.  
The gate fidelity is a real number that takes a value $0\le F\le 1$, and $F=1$ 
is achieved when there are no errors. Density plots of the gate fidelity 
of the target elementary pulse, BB1, CORPSE, and reduced CinBB 
are given in Fig.~\ref{fig:dfplot}. These figures clearly show that the 
reduced CCCP is simultaneously robust against both PLE and ORE while
others are not.

\section{Conclusion and Discussion}\label{sec:conc}
Composite pulses proposed so far suppress either PLE or ORE.  
As a straightforward extension of conventional composite pulses, 
we establish a general method 
to design single-qubit concatenated composite pulses (CCCPs),
which are robust against these two types of errors simultaneously.
Some composite pulses have residual-error-preserving (REP) property
that is a sufficient condition for them to be inner composite pulses 
of CCCPs.
There are various combinations of two composite pulses to design 
CCCPs. If the composite pulses chosen have one or more trivial pulses, 
the number of pulses and operation time cost are further reduced.  

In closing, we point out somewhat unexpected similarity between 
gate operations under error and those under noise~\cite{kv}.  
For example, similar to our no-go theorem for $N=2$ composite pulse, it has
been shown that dynamical decoupling pulses against any environment result in
the identity operation. Moreover, a balanced pair,
which is a two-pulse sequence
whose first order noise term is identical to that of the elementary pulse,
has been employed to implement a non-trivial pulse sequence robust
against noise, which is similar to our residual-error-preserving property.
Whether this similarity
is superficial or originates from deep connection between the two cases is
under investigation and will be published elsewhere.

\begin{acknowledgment}
We would like to thank Lorenza Viola for drawing our attention to 
their work~\cite{kv}.
This work is supported by ``Open Research Center'' Project for 
Private Universities; Matching Fund Subsidy from MEXT 
(Ministry of Education, Culture, Sports, Science and Technology), Japan. 
M.~B. is grateful to 
the Sasakawa Scientific Research Grant from The Japan Science Society.  
Y.~K. and M.~N. would like to thank partial supports of Grants-in-Aid for 
Scientific Research from the JSPS (Grant No. 23540470).  
\end{acknowledgment}



\end{document}